\documentclass[fleqn,10pt]{wlscirep}
\usepackage[utf8]{inputenc}
\usepackage[T1]{fontenc}
\usepackage{color}
\usepackage{subcaption}
\usepackage{rotating}
\usepackage{tabularx}

\title{Human-level COVID-19 Diagnosis from Low-dose CT Scans Using a Two-stage Time-distributed Capsule Network}

\author[1,7]{Parnian Afshar}
\author[2]{Moezedin Javad Rafiee, MD}
\author[1]{Farnoosh Naderkhani}
\author[3]{Shahin Heidarian}
\author[1]{Nastaran Enshaei}
\author[4]{Anastasia Oikonomou, MD}
\author[5]{Faranak Babaki Fard, MD}
\author[4]{Reut Anconina, MD}
\author[6]{Keyvan Farahani}
\author[7]{Konstantinos N. Plataniotis}
\author[1,*]{Arash Mohammadi}

\affil[1]{Concordia Institute for Information Systems Engineering (CIISE), Concordia University, Montreal, Canada}
\affil[2]{Department of Medicine and Diagnostic Radiology, McGill University Health Center-Research Institute, Montreal, QC, Canada}
\affil[3]{Department of Electrical and Computer Engineering, Concordia University, Montreal, QC, Canada}
\affil[4]{Department of Medical Imaging, Sunnybrook Health Sciences Centre, University of Toronto, Toronto, Canada}
\affil[5]{Faculty of Medicine, University of Montreal, Montreal, QC, Canada}
\affil[6]{Center for Biomedical Informatics and Information Technology, National Cancer Institute (NCI), Rockville, MD, USA.}
\affil[7]{Department of Electrical and Computer Engineering, University of Toronto, Toronto, Canada}

\affil[*]{Corresponding author: Arash Mohammadi (arash.mohammadi@concordia.ca)}

\begin{abstract}
Reverse transcription-polymerase chain reaction (RT-PCR) is currently the gold standard in COVID-19 diagnosis. It can, however, take days to provide the diagnosis, and false negative  rate is relatively high. Imaging, in particular chest computed tomography (CT), can assist with diagnosis and assessment of this disease. Nevertheless, it is shown that standard dose CT scan gives significant radiation burden to  patients, especially those in need of multiple scans. In this study, we consider low-dose and ultra-low-dose (LDCT and ULDCT) scan protocols  that reduce the radiation exposure close to that of a single X-Ray, while maintaining an acceptable resolution for diagnosis purposes. Since thoracic radiology expertise may not be widely available during the pandemic, we develop an Artificial Intelligence (AI)-based framework using a collected dataset of LDCT/ULDCT scans, to study the hypothesis that the AI model can provide human-level performance. The AI model uses a two stage capsule network architecture and can rapidly classify COVID-19, community acquired pneumonia (CAP), and normal cases, using LDCT/ULDCT scans. Based on a cross validation, the AI model achieves COVID-19 sensitivity of $89.5\%\pm 0.11$, CAP sensitivity of $95\%\pm 0.11$, normal cases sensitivity (specificity) of $85.7\%\pm 0.16$, and accuracy of $90\%\pm 0.06$. By incorporating clinical data (demographic and symptoms), the performance further improves to COVID-19 sensitivity of $94.3\%\pm 0.05$, CAP sensitivity of $96.7\%\pm 0.07$, normal cases sensitivity (specificity) of $91\%\pm 0.09$ , and accuracy of $94.1\%\pm 0.03$. The proposed AI model achieves human-level diagnosis based on the LDCT/ULDCT scans with reduced radiation exposure. We believe that the proposed AI model has the potential to assist the radiologists to accurately and promptly diagnose COVID-19 infection and help control the transmission chain during the pandemic.

\end{abstract}
\begin{document}

\flushbottom
\maketitle
\thispagestyle{empty}

\section*{Introduction}
Since the beginning of the coronavirus disease (COVID-19) outbreak in December 2019 in Wuhan, China, a global healthcare crisis has emerged~\cite{Wu}. Real-time reverse transcription-polymerase chain reaction (RT-PCR) is currently considered as the gold standard method in COVID-19 diagnosis. RT-PCR is, however, prone to a number of limitations, i.e.,  besides being time consuming, it is associated with high false-negative rate in different clinical samples~\cite{Fang}. Due to high sensitivity and rapid access, chest computed tomography (CT) scan has been the main imaging modality for diagnosis, prognostic assessment, and detection of complications of COVID-19~\cite{Salehi,Rafiee}. The most common manifestations of COVID-19 pneumonia in chest CT scan are multifocal ground-glass opacities (GGO) with or without consolidative areas, predominantly having peripheral, lower-lobes, and posterior anatomic distribution~\cite{Salehi,Rafiee}.  CT scan can contribute to assessing the complications, extent of COVID-19 involvement, and risk of intensive care unit (ICU) admission.

The main concern of widespread use of CT scan as a screening tool for suspected patients during the outbreak is the radiation exposure. In some scenarios, severely symptomatic patients will need multiple chest CT scans during the course of their disease. The cumulative effect of these multiple exposures can significantly increase the radiation dose. Studies~\cite{Miglioretti} have shown that the projected radiation to body organs during chest CT scan is highest in thyroid, lung, breast, and esophagus. Due to their longer life expectancy, higher dose-effective breast tissue and cell proliferation~\cite{Fearon,Pearce}, children and young women are more vulnerable to radiation exposure damage with increased risk of radiation-following malignancy.  As low as reasonably achievable (ALARA)~\cite{Bell} rule states that whenever radiation is expected, the exposure should be kept at the minimum achievable level such that the resulting scan still provides reasonable resolution.

Diagnostic accuracy of Low and Ultra-low-dose CT scan (LDCT and ULDCT) in detection and follow-up of pulmonary nodule and other lung pathologies has been previously established~\cite{Taekker}. The radiation dose associated with standard chest CT is estimated at 7 mSv, which is reduced to 1-1.5 mSv with  LDCT methods and as low as 0.3 mSv with ULDCT ones. The advantage of the low dose protocols is the reduction of radiation dose by more than 80\%. Recent studies~\cite{Sakane} have shown that DNA double-strand breaks and chromosome damage increased in patients undergoing a standard-dose CT scan while no effect on human DNA was demonstrated in patients undergoing low-dose CT scan. LDCT and ULDCT have shown significant accuracy in the detection of GGOs and consolidation in patients with pneumonia~\cite{Park}. Since GGO and consolidation are the most common CT findings of COVID-19, recently, replacing standard CT scan with LDCT and ULDCT has been recommended~\cite{Schulze-Hagen} as a solution to decrease radiation exposure in COVID-19 patients. In a retrospective study~\cite{Dangis}, LDCT with iterative reconstruction (IR) demonstrated sensitivity, specificity, positive predictive value, negative predictive value, and accuracy of about 90\% in the diagnosis of COVID-19. In conjunction with other clinical findings,  LDCT and ULDCT can potentially replace standard-dose  for the evaluation of patients, in particular pregnant and  young women, and pediatric populations, to decrease radiation exposure~\cite{Tofighi}.

While rapid detection of positive COVID-19 cases is of utmost importance, physicians and healthcare personnel are overwhelmed with increasing number of patients in need of immediate care and treatment. In other words, expert thoracic radiologists may not be available at all times to diagnose positive cases in a timely fashion, leading to not only delays  in treatment, but also further transmission of the virus by patients who are not promptly isolated. Recent studies~\cite{Ardila, Espinoza} have demonstrated the capabilities of artificial intelligence (AI) models in achieving human-level performance in detection of lung nodules from LDCT. Motivated by this, in this study we aim at developing an AI model based on  deep learning architectures to distinguish COVID-19 from community acquired pneumonia (CAP) and normal cases, using LDCT and ULDCT. To the best of our knowledge, Reference~\cite{Shiri} is the only study considering LDCT in AI-based COVID-19 analysis, by simulating standard dose scans from low dose ones. The aforementioned study, however, uses synthesized data, i.e., it does not use real LDCT/ULDCT data from COVID-19 individuals, and does not deal with the disease diagnosis, which is the main focus of our research. We hypothesize that AI can achieve a human-level performance in diagnosing COVID-19 based on LDCT and ULDCT scans.

\begin{figure}[ht]
\centering
\includegraphics[width=\linewidth]{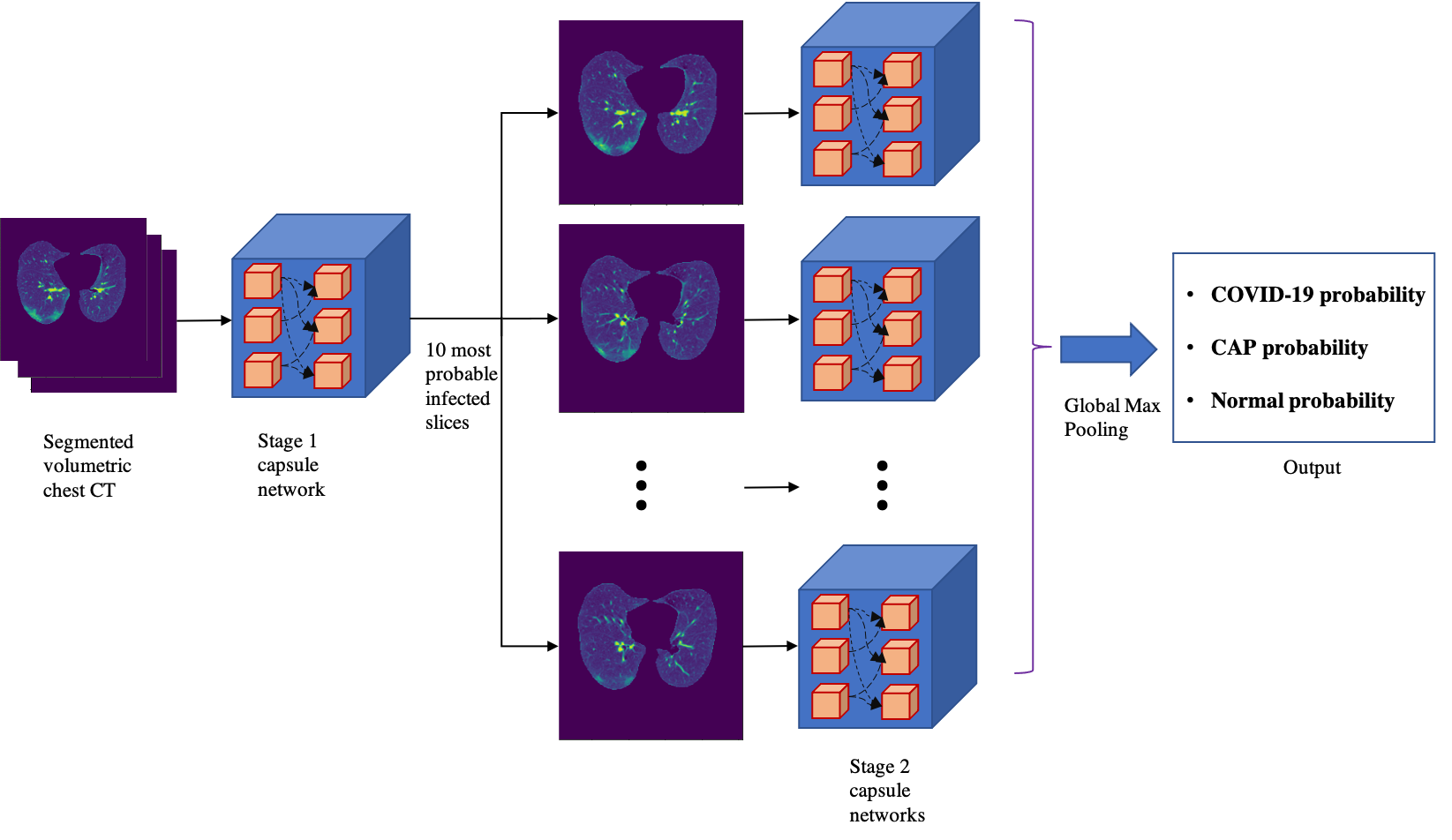}
\caption{\small The proposed 2 stage deep learning model for COVID-19 diagnosis using LDCT/ULDCT. At the first stage, CT slices go through a capsule network, one by one, to detect those with evidence of infection. At the second stage, 10 most probable slices with infection detected in the previous stage go through a time-distributed capsule network, output of which determines the probability of COVID-19, CAP, and normal, after applying a global max pooling.}
\label{fig:framework}
\end{figure}

\begin{figure}[ht]
\centering
\includegraphics[width=\linewidth]{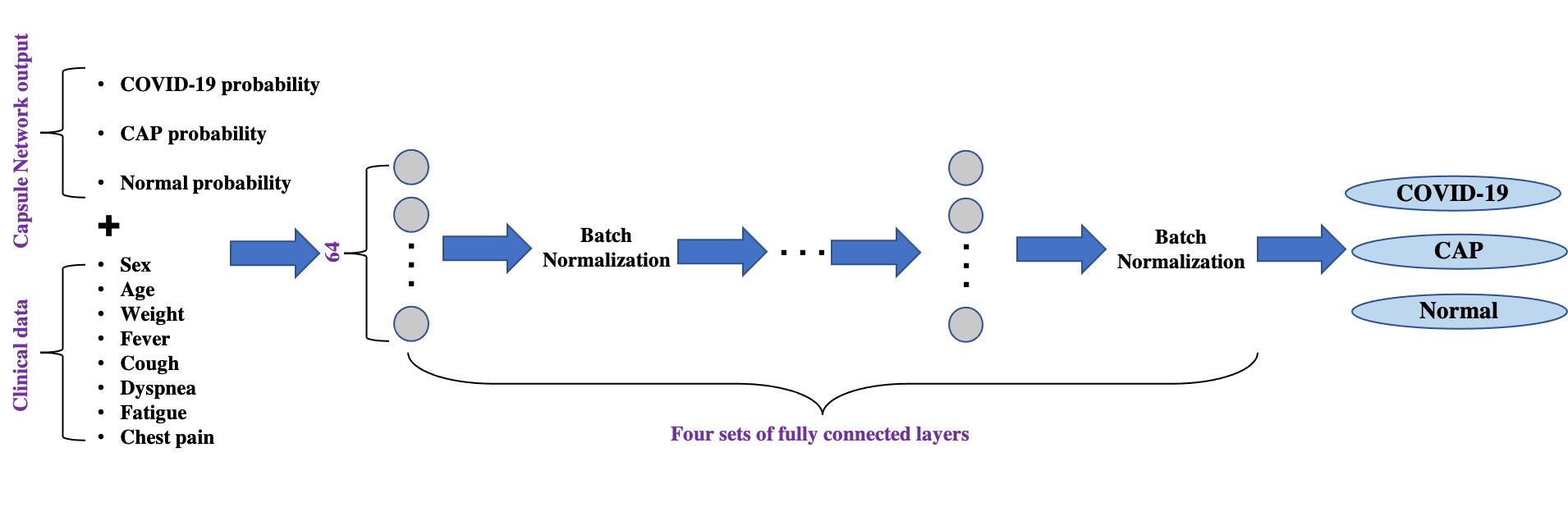}
\caption{\small The MLP model combining the output of the two-stage deep learning model with the clinical data. Clinical data includes demographic characteristics and 5 common COVID-19 and CAP symptoms. Four sets of fully connected layers determine the final output.}
\label{fig:mlp}
\end{figure}

We developed a two-stage deep learning model, shown in Fig.~\ref{fig:framework}, built upon the capsule network architecture~\cite{Sabour, Afshar3DMCN}, which takes segmented lung regions as inputs. The first stage of the proposed model is a capsule network responsible of detecting slices with evidence of infection (caused by COVID-19 or CAP). Following a prior research~\cite{Mei}, 10 most probable slices with infection, along with their infectious probability, are provided as input to the second stage, which analyzes all the candidates in parallel, making the decision by aggregating the outcome of each single candidate. The final output is the probability of the underlying case representing  COVID-19, CAP, or normal classes. While the first stage model is trained on slice-level labels provided by an experienced radiologist, the second stage is trained on patient-level labels. Expanding the proposed deep learning model, we incorporated clinical data, i.e., the output of the deep learning model is merged with clinical data and fed to a multi-layer perceptron (MLP), shown in Fig.~\ref{fig:mlp}, to make the final diagnosis.

\section*{Results}
\begin{table}[t]
\centering
\caption{\small Characteristics of the collected dataset and CAP cases adopted from Reference~\cite{afshar:nsd}. In total, the study includes 220 patients}. SD stands for standard deviation, M stands for male and F stands for female.
\label{tab:data}
\begin{tabularx}{\textwidth}{X|X|X|X|X|X|X|}
\cline{2-7}
 & \multicolumn{1}{|c|}{\textbf{COVID-19}}
 & \multicolumn{1}{|c|}{\textbf{CAP}}
 & \multicolumn{1}{|c|}{\textbf{Normal}}
  & \multicolumn{1}{|X|}{\textbf{P-value: COVID-19  vs. rest}}
  & \multicolumn{1}{|X|}{\textbf{P-value: CAP  vs. rest}}
  & \multicolumn{1}{|X|}{\textbf{P-value: Normal  vs. rest}}\\
\hline
 \multicolumn{1}{|c|}{Sex} & M: 58.6\% ~~~F: 41.4\%& M: 58\% ~~~~~~F: 42\% & M: 39.3\% ~~~F: 60.7\%&0.7386  & 0.1848  & 0.0314 \\
\hline
\multicolumn{1}{|c|}{Age in Years (Mean $\pm$ SD)} & $49.53\pm 15.5$ & $57.78\pm 21.94$ & $40.18 \pm 15.37$ &0.7283 & 0.0003 & 0.0002 \\
\hline
\multicolumn{1}{|c|}{Weight in Kg (Mean $\pm$ SD)} & $80.75 \pm 14.84$ & $67.38 \pm 12.96$ & $75.91 \pm 14.52$ &0.0001 & 0.0000  & 0.6881\\
\hline
\multicolumn{1}{|c|}{Dyspnea} & $26.9\%$ & $18\%$ & $45\%$  &0.8932 & 0.0091 & 0.0425\\
\hline
\multicolumn{1}{|c|}{Cough} & $31.7\%$ & $53\%$ & $33.93\%$ &0.1480& 0.0160 & 0.4916 \\
\hline
\multicolumn{1}{|c|}{Fever} & $14.4\%$ & $36\%$ & $9\%$ &0.5130 & 0.0242& 0.0589 \\
\hline
\multicolumn{1}{|c|}{Chest Pain} & $7\%$ & $0\%$ & $10.7\%$ &0.7571 & 0.9999 & 0.6439\\
\hline
\multicolumn{1}{|c|}{Fatigue} & $10.5\%$ & $0\%$ & $1.7\%$ &0.0107 & 0.9999 & 0.2681\\
\hline
\end{tabularx}
\vspace{-.2in}
\end{table}

We collected a dataset of LDCT and ULDCT scans of 104 COVID-19 positive cases, and 56 normal cases, collected in October 2020, December 2020, and January 2021, Babak Imaging Center, Tehran, Iran. Diagnosis of 36.5\% of the COVID-19 cases (38 cases)  is confirmed with the RT-PCR test. The rest are specified by taking the consensus between 3 experienced thoracic radiologists (M.J.R., F.B.F., and A.O.), who labeled the dataset by taking the imaging findings, clinical characteristics (symptoms and history), and epidemiology into account. The three radiologists reached an agreement of 95.6\%. They also scored the severity of the COVID-19 cases between 1 and 4, based on the percentage of the lung involvement. Four positive COVID-19 cases do not reveal any related imaging findings. As we did not have access to LDCT scans of CAP patients, we combined this dataset with 60 standard-dose volumetric CT scans~\cite{afshar:nsd}. Therefore, we ended up with a total of 220 patients.
The dataset characteristics are shown in Table~\ref{tab:data}. P-values are obtained using logistic regression, by considering three binary scenarios of COVID-19 versus CAP and normal, CAP versus COVID-19 and normal, and normal versus COVID-19 and CAP. For ease of access, here we translate the first row of the table: "58.6\% of the COVID-19 patients are men, 58\% of the CAP cases are men, 39.3\% of the normal cases are men, Sex has a p-value of 0.7386 when distinguishing COVID-19 from the other two classes, it has a p-value of 0.1848 when distinguishing CAP from the other two classes, and a p-value of 0.0314 when distinguishing normal from other two classes".
Finally, a fourth experienced thoracic radiologist (R.A.), blind to the ground-truth, labeled the collected dataset  to compare the performance of the AI model with a human expert. The radiologist was first provided with only the CT scans, and then the clinical data.

To decrease bias towards a specific test set, we adopted a 10-fold cross validation approach to assess the performance of the radiologist and the AI model, based on two scenarios of using CT scans only, and incorporating the clinical data. The dataset is randomly split into 10 equal size test sets, leading to 10 sets each including 22 cases. We made sure that each set contained 10\% of the COVID-19, CAP, and normal cases, leading to 10 or 11 COVID-19, 6 CAP, and 5 or 6 normal cases in each test set. The AI model is trained 10 times, setting one of the test sets aside and using the rest for training.  Averaging over the 10 folds, the slice-level classifier in the first stage achieved  accuracy of $89.88\%$, sensitivity of $88.24\%$, and specificity of $92.01\%$, in detecting the slices with infection.

\begin{sidewaystable}
\centering
\caption{\small Performance of the AI model and the radiologist blind to the labels, using only CT scans.}
\label{tab:res1}
\begin{tabular}{c|c|c|c|c|c|c|c|c|c|c|}
\cline{2-9}
 & \multicolumn{2}{|c|}{\textbf{COVID-19 Sensitivity}}
 & \multicolumn{2}{|c|}{\textbf{CAP Sensitivity}}
 & \multicolumn{2}{|c|}{\textbf{Normal Sensitivity}}
 & \multicolumn{2}{|c|}{\textbf{Accuracy}}\\
\hline
 \multicolumn{1}{|c|}{Fold} & AI  & Radiologist & AI & Radiologist  & AI &  Radiologist & AI &  Radiologist  & Accuracy P-value &COVID-19 AUC \\
\hline
\multicolumn{1}{|c|}{1} & $\frac{10}{11}$ & $\frac{10}{11}$ & $\frac{6}{6}$ & $\frac{5}{6}$ & $\frac{4}{5}$  & $\frac{5}{5}$ & $\frac{20}{22}$ & $\frac{20}{22}$ & $1$ & $0.95$\\[5pt]
\hline
\multicolumn{1}{|c|}{2} & $\frac{10}{10}$ & $\frac{10}{10}$ & $\frac{6}{6}$ & $\frac{6}{6}$ & $\frac{6}{6}$  & $\frac{6}{6}$ & $\frac{22}{22}$ & $\frac{22}{22}$ & $1$ & $1$\\[5pt]
\hline
\multicolumn{1}{|c|}{3} & $\frac{10}{11}$ & $\frac{9}{11}$ & $\frac{6}{6}$ & $\frac{4}{6}$ & $\frac{3}{5}$  & $\frac{5}{5}$ & $\frac{19}{22}$ & $\frac{18}{22}$ & $1$ & $0.91$\\[5pt]
\hline
\multicolumn{1}{|c|}{4} & $\frac{10}{10}$ & $\frac{10}{10}$ & $\frac{4}{6}$ & $\frac{6}{6}$ & $\frac{5}{6}$  & $\frac{6}{6}$ & $\frac{19}{22}$ & $\frac{22}{22}$ & $0.25$ & $0.99$\\[5pt]
\hline
\multicolumn{1}{|c|}{5} & $\frac{8}{11}$ & $\frac{10}{11}$ & $\frac{5}{6}$ & $\frac{5}{6}$ & $\frac{5}{5}$  & $\frac{5}{5}$ & $\frac{18}{22}$ & $\frac{20}{22}$ & $0.5$ & $0.9$\\[5pt]
\hline
\multicolumn{1}{|c|}{6} & $\frac{10}{11}$ & $\frac{10}{11}$ & $\frac{6}{6}$ & $\frac{5}{6}$ & $\frac{5}{5}$  & $\frac{5}{5}$ & $\frac{21}{22}$ & $\frac{20}{22}$ & $1$ & $0.98$\\[5pt]
\hline
\multicolumn{1}{|c|}{7} & $\frac{8}{10}$ & $\frac{10}{10}$ & $\frac{6}{6}$ & $\frac{5}{6}$ & $\frac{5}{6}$  & $\frac{6}{6}$ & $\frac{19}{22}$ & $\frac{21}{22}$ & $0.625$ & $0.96$\\[5pt]
\hline
\multicolumn{1}{|c|}{8} & $\frac{7}{10}$ & $\frac{8}{10}$ & $\frac{6}{6}$ & $\frac{5}{6}$ & $\frac{5}{6}$  & $\frac{6}{6}$ & $\frac{18}{22}$ & $\frac{19}{22}$ & $1$ & $0.95$\\[5pt]
\hline
\multicolumn{1}{|c|}{9} & $\frac{10}{10}$ & $\frac{6}{10}$ & $\frac{6}{6}$ & $\frac{6}{6}$ & $\frac{5}{6}$  & $\frac{6}{6}$ & $\frac{21}{22}$ & $\frac{18}{22}$ & $0.375$ & $1$\\[5pt]
\hline
\multicolumn{1}{|c|}{10} & $\frac{10}{10}$ & $\frac{10}{10}$ & $\frac{6}{6}$ & $\frac{6}{6}$ & $\frac{5}{6}$  & $\frac{6}{6}$ & $\frac{21}{22}$ & $\frac{22}{22}$ & $1$ & $1$\\[5pt]
\hline
\multicolumn{1}{|c|}{Total} & $89.5\%\pm 0.11$ & $89.4\%\pm 0.12$ & $95\%\pm 0.11$ & $88.33\%\pm 0.11$ & $85.7\%\pm 0.16$  & $100\%$ & $90\%\pm 0.06$ & $91.8\%\pm 0.07$ & - & $0.96\pm 0.03$\\[5pt]
\hline
\end{tabular}
\vspace{-.2in}
\end{sidewaystable}

\begin{figure}[ht]
\centering
\includegraphics[width=0.5\linewidth]{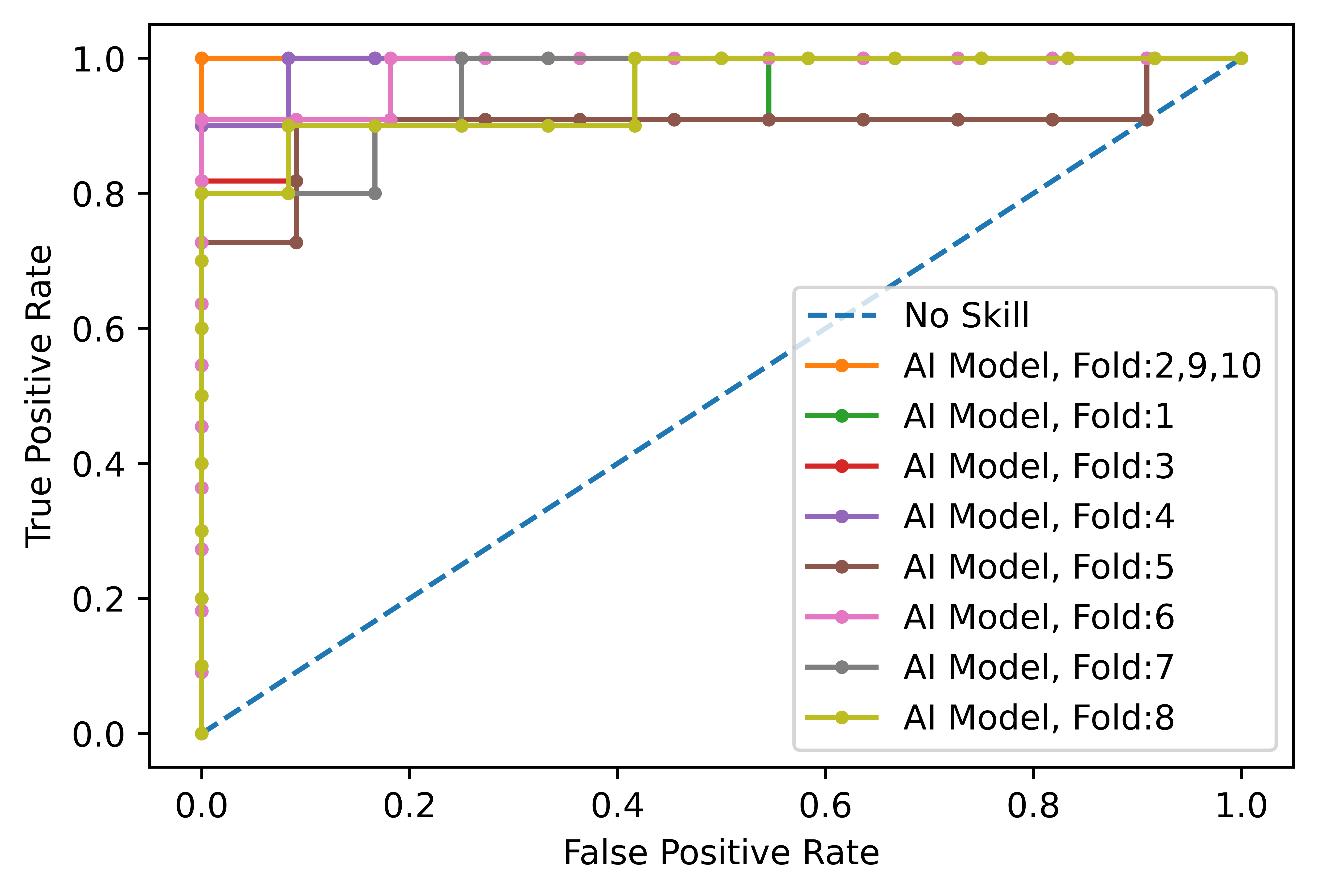}
\caption{\small ROC curve for COVID-19 diagnosis (vs CAP and normal) using the proposed deep learning model and CT scans only.}
\label{fig:ROC_CT}
\end{figure}

Using only CT scans, we evaluated the developed deep learning model and compared it with the fourth thoracic radiologist, as shown in Table~\ref{tab:res1}. Averaging over all the 10 folds, AI model achieves COVID-19 sensitivity of $89.5\%\pm 0.11$, CAP sensitivity of $95\%\pm 0.11$, normal sensitivity (specificity) of $85.7\%\pm 0.16$, and accuracy of $90\%\pm 0.06$. The radiologist, on the other hand, achieves COVID-19 sensitivity of $89.4\%\pm 0.12$, CAP sensitivity of $88.33\%\pm 0.11$, normal sensitivity (specificity) of $100\%$, and accuracy of $91.8\%\pm 0.07$. We tested the hypothesis of the AI model and radiologist having the same performance, in term of accuracy, using a McNemar~\cite{McNemar} test with the significance level of 0.05, leading to P-values over the significance level for all the 10 folds. The lower specificity of the AI model conforms the non-specific COVID-19 findings~\cite{Prokop}. COVID-19 sensitivity versus one minus specificity is plotted in the receiver operating characteristics (ROC) curve, shown in Fig.~\ref{fig:ROC_CT}. Area under the curve (AUC) is $0.96\pm 0.03$.

\begin{figure}[ht]
\centering
\includegraphics[width=\linewidth]{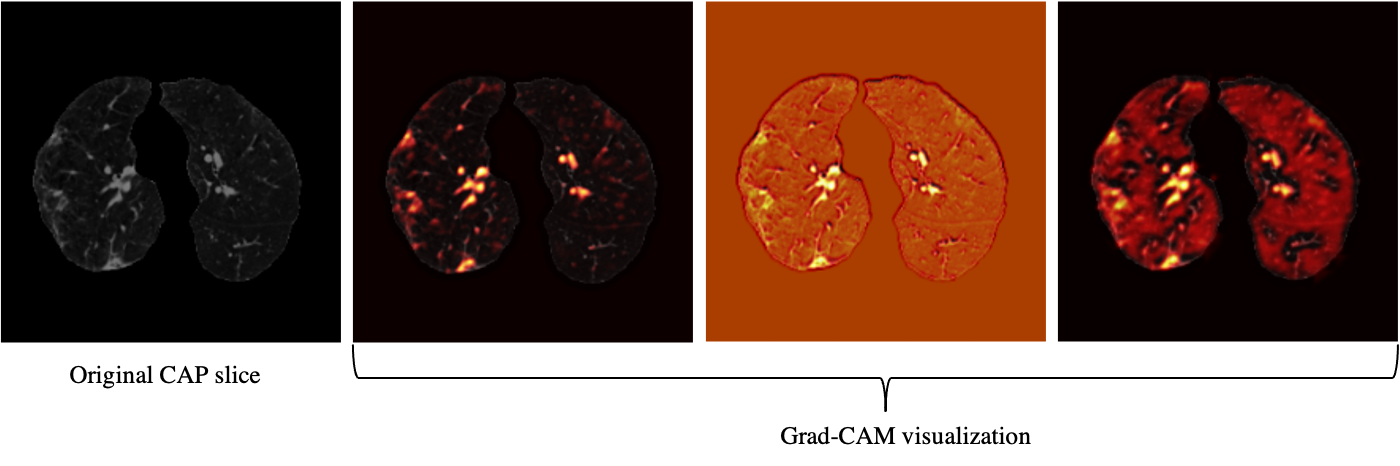}
\caption{\small Grad-CAM visualization of one CAP slice. This figure shows that the proposed AI model is paying attention to relevant locations of the image.}
\label{fig:Grad}
\end{figure}

\begin{figure}[ht]
\centering
\includegraphics[width=\linewidth]{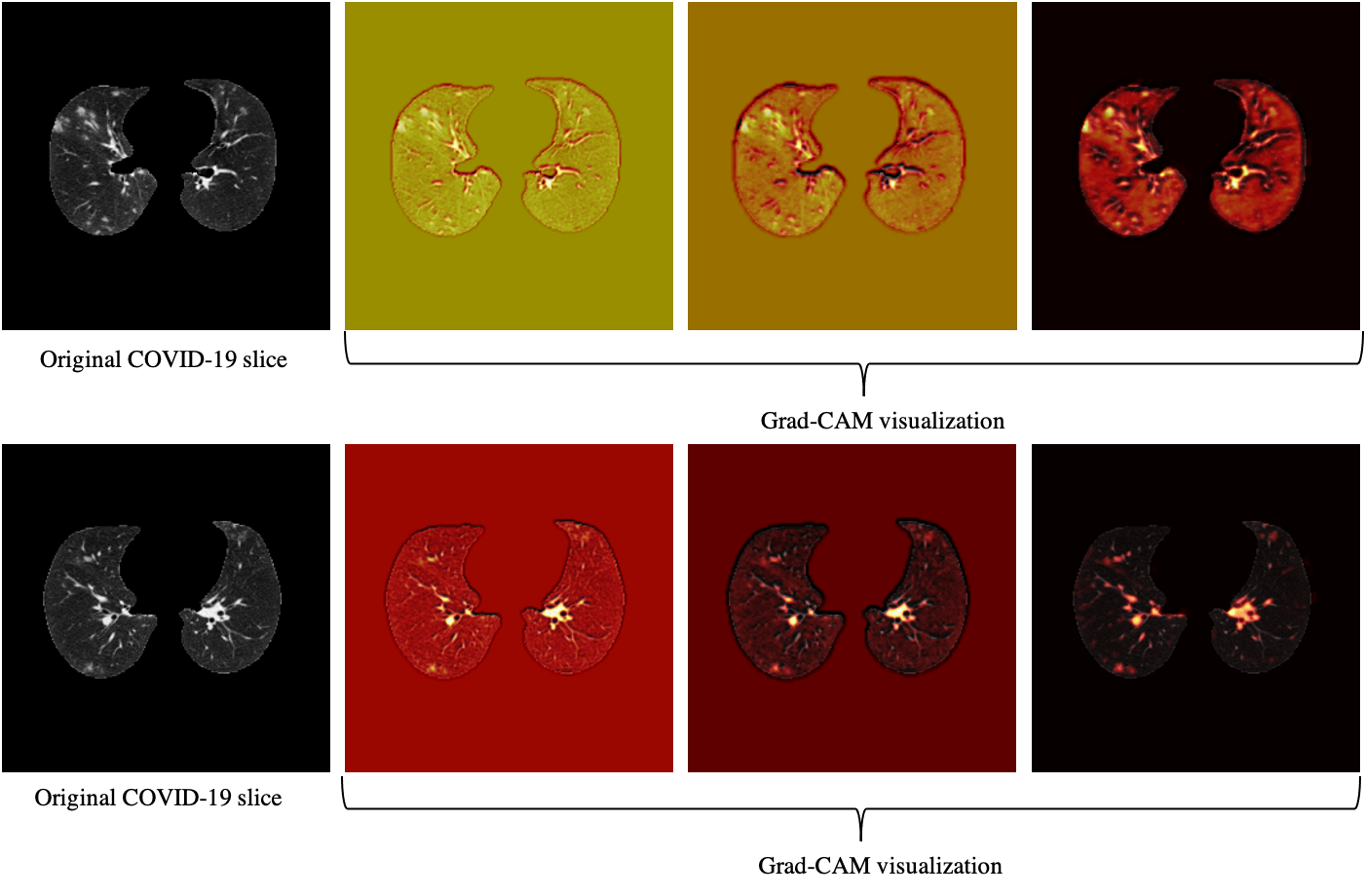}
\caption{\small Grad-CAM visualization of two COVID-19 slices. This figure shows that the proposed AI model is paying attention to relevant locations of the image.}
\label{fig:Grad2}
\end{figure}

Based on the CT scans only, we analyzed the misclassified COVID-19 cases through all folds (11 cases in total), and studied their relation with the disease severity, coming to the conclusion that 4 out of 11 cases,  did not have any related imaging findings, 5 were scored 1 by the three radiologists, one was scored 2, and only one case was scored at 3, which means the developed model is less likely to misclassify severe cases. Neither the developed model nor the experienced radiologist was able to detect the 4 COVID-19 cases without imaging findings, using CT scans only. Furthermore, since the CAP patients come from a different cohort and scanned with a standard dose, we visualized the model's output for CAP cases, one of which is shown in Fig.~\ref{fig:Grad}, using Grad-CAM localization technique. This figure shows that the model is paying more attention to disease-related regions of the image, rather than dose-related ones. We performed the same localization technique on two slices with infection of the same COVID-19 patient, shown in Fig.~\ref{fig:Grad2}.

\begin{sidewaystable}
\centering
\caption{\small Performance of the AI model and the radiologist blind to the labels, using both CT scans and clinical data.}
\label{tab:res2}
\begin{tabular}{c|c|c|c|c|c|c|c|c|c|c|}
\cline{2-9}
 & \multicolumn{2}{|c|}{\textbf{COVID-19 Sensitivity}}
 & \multicolumn{2}{|c|}{\textbf{CAP Sensitivity}}
 & \multicolumn{2}{|c|}{\textbf{Normal Sensitivity}}
 & \multicolumn{2}{|c|}{\textbf{Accuracy}}\\
\hline
 \multicolumn{1}{|c|}{Fold} & AI  & Radiologist & AI & Radiologist  & AI &  Radiologist & AI &  Radiologist  & Accuracy P-value &COVID-19 AUC \\
\hline
\multicolumn{1}{|c|}{1} & $\frac{11}{11}$ & $\frac{10}{11}$ & $\frac{5}{6}$ & $\frac{6}{6}$ & $\frac{5}{5}$  & $\frac{5}{5}$ & $\frac{21}{22}$ & $\frac{21}{22}$ & $1$ & $0.99$\\[5pt]
\hline
\multicolumn{1}{|c|}{2} & $\frac{10}{10}$ & $\frac{10}{10}$ & $\frac{6}{6}$ & $\frac{6}{6}$ & $\frac{6}{6}$  & $\frac{6}{6}$ & $\frac{22}{22}$ & $\frac{22}{22}$ & $1$ & $1$\\[5pt]
\hline
\multicolumn{1}{|c|}{3} & $\frac{10}{11}$ & $\frac{10}{11}$ & $\frac{6}{6}$ & $\frac{6}{6}$ & $\frac{4}{5}$  & $\frac{5}{5}$ & $\frac{20}{22}$ & $\frac{21}{22}$ & $1$ & $0.91$\\[5pt]
\hline
\multicolumn{1}{|c|}{4} & $\frac{9}{10}$ & $\frac{10}{10}$ & $\frac{5}{6}$ & $\frac{6}{6}$ & $\frac{6}{6}$  & $\frac{6}{6}$ & $\frac{20}{22}$ & $\frac{22}{22}$ & $0.5$ & $0.98$\\[5pt]
\hline
\multicolumn{1}{|c|}{5} & $\frac{10}{11}$ & $\frac{10}{11}$ & $\frac{6}{6}$ & $\frac{5}{6}$ & $\frac{5}{5}$  & $\frac{5}{5}$ & $\frac{21}{22}$ & $\frac{20}{22}$ & $1$ & $0.97$\\[5pt]
\hline
\multicolumn{1}{|c|}{6} & $\frac{10}{11}$ & $\frac{10}{11}$ & $\frac{6}{6}$ & $\frac{5}{6}$ & $\frac{4}{5}$  & $\frac{5}{5}$ & $\frac{20}{22}$ & $\frac{20}{22}$ & $1$ & $0.99$\\[5pt]
\hline
\multicolumn{1}{|c|}{7} & $\frac{9}{10}$ & $\frac{10}{10}$ & $\frac{6}{6}$ & $\frac{5}{6}$ & $\frac{5}{6}$  & $\frac{6}{6}$ & $\frac{20}{22}$ & $\frac{21}{22}$ & $1$ & $1$\\[5pt]
\hline
\multicolumn{1}{|c|}{8} & $\frac{9}{10}$ & $\frac{10}{10}$ & $\frac{6}{6}$ & $\frac{5}{6}$ & $\frac{6}{6}$  & $\frac{6}{6}$ & $\frac{21}{22}$ & $\frac{21}{22}$ & $1$ & $0.98$\\[5pt]
\hline
\multicolumn{1}{|c|}{9} & $\frac{10}{10}$ & $\frac{8}{10}$ & $\frac{6}{6}$ & $\frac{6}{6}$ & $\frac{5}{6}$  & $\frac{6}{6}$ & $\frac{21}{22}$ & $\frac{20}{22}$ & $1$ & $0.95$\\[5pt]
\hline
\multicolumn{1}{|c|}{10} & $\frac{10}{10}$ & $\frac{10}{10}$ & $\frac{6}{6}$ & $\frac{6}{6}$ & $\frac{5}{6}$  & $\frac{6}{6}$ & $\frac{21}{22}$ & $\frac{22}{22}$ & $1$ & $0.99$\\[5pt]
\hline
\multicolumn{1}{|c|}{Total} & $94.3\%\pm 0.05$ & $94.4\%\pm 0.05$ & $96.7\%\pm 0.07$ & $93.3\%\pm 0.08$ & $91\%\pm 0.09$  & $100\%$ & $94.1\%\pm 0.03$ & $95.4\%\pm 0.03$ & - & $0.96\pm 0.03$\\
\hline
\end{tabular}
\vspace{-.2in}
\end{sidewaystable}

\begin{figure}[ht]
\centering
\includegraphics[width=0.5\linewidth]{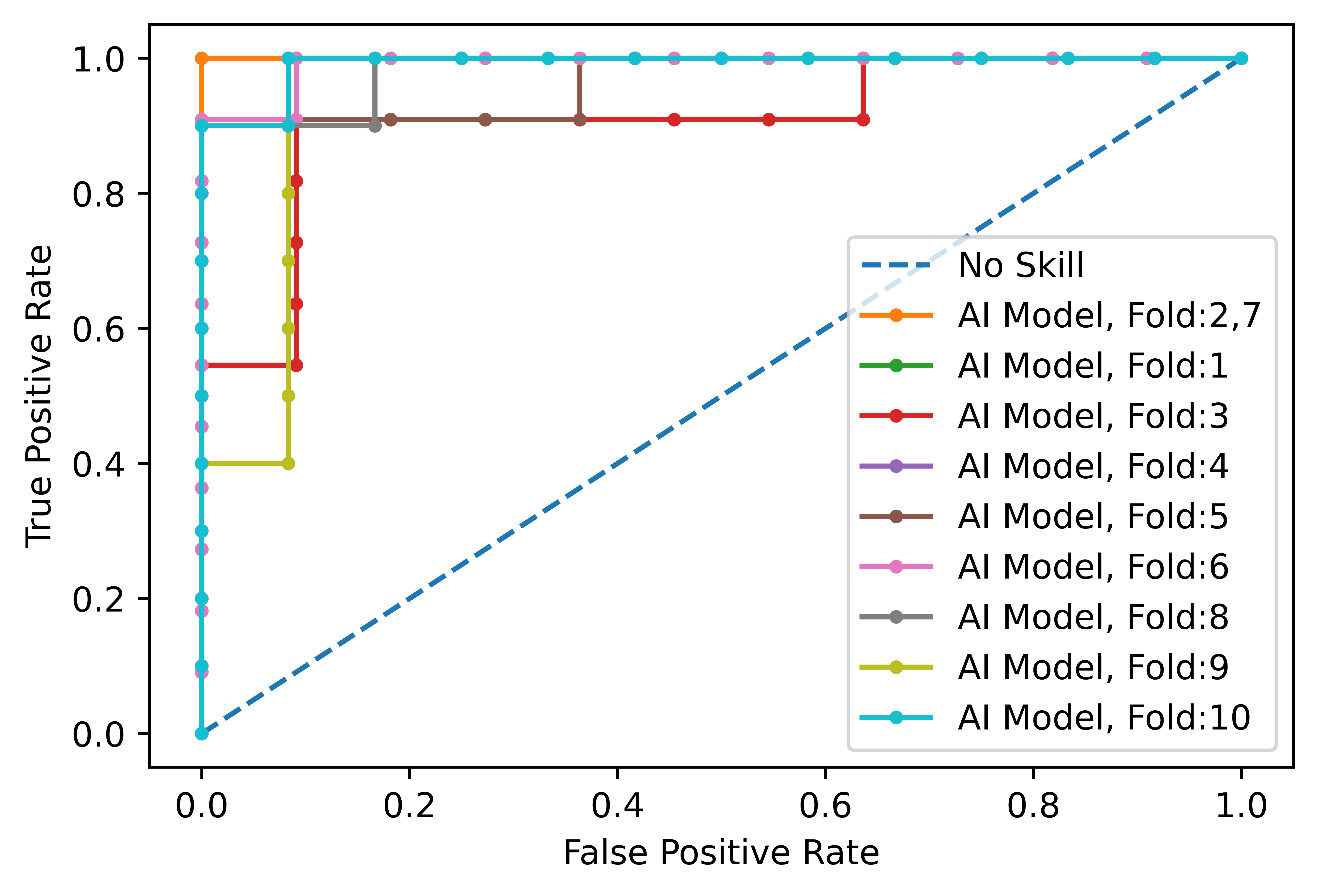}
\caption{\small ROC curve for COVID-19 diagnosis (vs CAP and normal) using the proposed deep learning model and both CT scans and clinical data.}
\label{fig:ROC_Clinical}
\end{figure}

Using both CT scans and clinical data, we evaluated the developed deep learning model and compared it with the radiologist, as shown in Table~\ref{tab:res2}. Averaging over all the 10 folds, AI model achieves  COVID-19 sensitivity of $94.3\%\pm 0.05$, CAP sensitivity of $96.7\%\pm 0.07$, normal sensitivity (specificity) of $91\%\pm 0.09$ , and accuracy of $94.1\%\pm 0.03$. The radiologist, on the other hand, achieves  COVID-19 sensitivity of $94.4\%\pm 0.05$, CAP sensitivity of $93.3\%\pm 0.08$, normal sensitivity (specificity) of $100\%$, and accuracy of $95.4\%\pm 0.03$. We tested the hypothesis of the AI model and radiologist having the same performance, using LDCT and clinical data, in terms of accuracy, leading to P-values over the significance level for all the 10 folds.  COVID-19 sensitivity versus one minus specificity is plotted in the receiver operating characteristics (ROC) curve, shown in Fig.~\ref{fig:ROC_Clinical}. Area under the curve (AUC) is $0.96\pm 0.03$.

Based on using both CT scans and clinical data, we analyzed the misclassified COVID-19 cases through all folds (6 cases in total), and studied their relation with the disease severity, coming to the conclusion that 3 out of 6 cases,  did not have any related imaging findings, one was scored 1 by the three radiologists, and two cases were scored at 3. Incorporating the clinical data, the AI model can detect one of the four positive COVID-19 cases, without having related imaging findings, whereas the radiologist did not detect any of them.

Finally, we tested the developed AI model, incorporating LDCTs and clinical data, on an extra set of 100 positive COVID-19 patients, whose diagnosis are confirmed with RT-PCR test and are collected in a different time interval (narrow validation). These patients were not included in any of the 10 folds and are completely unseen to the model and radiologist. While 68 out of 100 cases have imaging findings, 32 do not reveal any related manifestations. Male cases constitute 53\% of the total cases, and age average is $46.16$ with a standard deviation of $14.07$. The AI model correctly identifies all the 68 positive cases having imaging findings, whereas it detects only 3 of those not having related findings. Radiologist, on the other hand, correctly classifies 64 out of 68 patients having imaging findings as COVID-19 and classifies 4 as CAP. None of the cases without imaging findings are identified by the radiologist. The p-value between the AI model and radiologist's sensitivity is 0.01.

\section*{Discussion}
\begin{figure}[ht]
\centering
\includegraphics[width=0.7\linewidth]{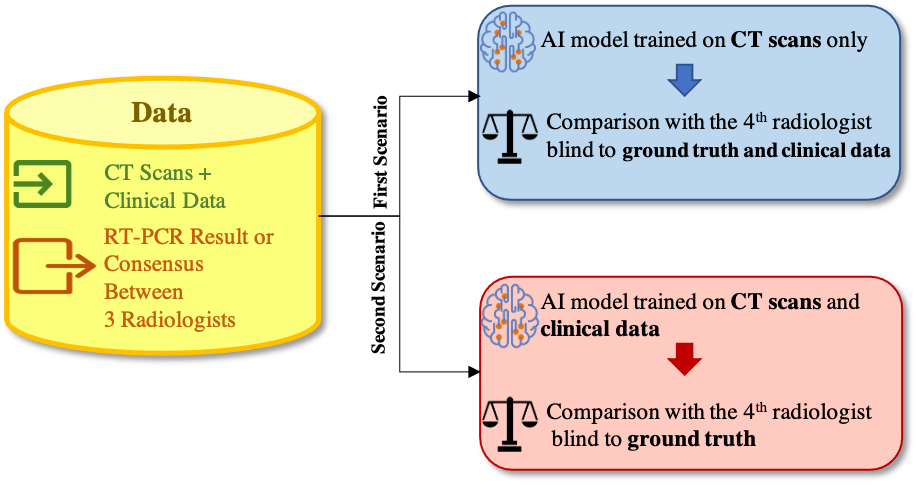}
\caption{\small The dataset consists of LDCTs and ULDCTs accompanied with clinical data and ground truth. Two scenarios are considered. First, only images are used for training and comparison with the radiologist. Second, both images and clinical data are utilized.}
\label{fig:scenario}
\end{figure}
Although LDCT and ULDCT can reveal COVID-19 related findings and reduce the potential radiation-related harms, an accurate diagnosis requires full investigation by radiologists, which may not be possible during the outbreak. Based on our experiments, the proposed capsule network-based AI model has the potential to rapidly distinguish COVID-19 cases from CAP and normal ones with a human-level performance using LDCT and ULDCT, having a radiation dose of a single X-ray image. In other words, with minimal radiation, the developed AI system can assist the radiologists and contribute to controlling the chain of COVID-19 transmission. 

To validate the proposed AI model, we considered two scenarios, as shown in Fig.~\ref{fig:scenario}. In the first scenario, the AI model is fed with only the images and compared with the radiologist blind to both ground truth and clinical data. Although this strategy does not follow routine clinical practice, the goal was to investigate the diagnostic potentials of the images without incorporating clinical data. In the second scenario, both images and clinical data are fed to the AI model which is accordingly compared to the radiologist blind only to the ground truth. We showed that by incorporating the clinical data, COVID-19 sensitivity increases by 4.8\%, CAP sensitivity increases by 1.7\%, and normal sensitivity and accuracy increase by 5.3\% and 4.1\%, respectively.

It is worth noting that although the incorporated CAP cases are extracted from a different cohort, the motivation is to investigate the capability of the AI model to distinguish COVID-19 from CAP, as these two have overlapping chest CT findings~\cite{Altmayer}. The fact that the CAP cases are screened using a standard-dose CT rather than LDCT and ULDCT can be considered as a limitation of our study, since low-dose screening is associated with less detection capabilities~\cite{Jonas}. Nevertheless it is a common practice to construct datasets of CT scans acquired using different radiation doses~\cite{Armato}, in order to develop models generalized on larger datasets. Furthermore, although our COVID-19 and normal cases are scanned with either LDCT or ULDCT, the image quality of the latter can be counterbalanced using reconstruction techniques~\cite{Finance}.

Our study has some other limitations. First, the dataset is collected from a single centre, and experiments are required to verify its performance on data from external institutes, as it is critical to investigate if the model generalizes to diverse population~\cite{WuE,Kleppe}. Vulnerability to data shifts, and bias against underrepresented population~\cite{WuE} are also crucial to address before the AI model can be put into practice.
It is worth mentioning that as the extra set of 100 positive COVID-19 patients are collected in a disjoint time interval from the original set, it can act as a narrow validation~\cite{Kleppe}. It is, however, collected from the same institute and thus does not account for broad validation. It is also of high interest to explore domain validation for COVID-19 diagnosis, where test set comes from different variants.
Second, the sample size is relatively small. Verifying the model's performance on larger multi-centre datasets is the goal of our upcoming studies.  The capsule network used in our study, is capable of handling small datasets compared to conventional models and due to fewer trainable parameters it is less prone to over-fitting, however, larger datasets can still improve the performance of the model. We also aim at expanding the proposed AI model to predict the disease severity besides the diagnosis. Moreover, although as shown in Figs.~\ref{fig:Grad}  and~\ref{fig:Grad2} visualization of the AI model's output shows it is paying attention to relevant regions, more research is required to increase its  explainability. Low performance on COVID-19 cases without imaging finding is another limitation of the developed model.

In conclusion, we believe the developed AI model achieves human-level performance by incorporating LDCT/ULDCT and clinical data, having the advantage of reducing the risks related to radiation exposure.  This model can act as a decision support system for radiologists and help with controlling the transmission chain. As our developed AI model is not intended to be a primary diagnostic tool, we aim at testing the model alongside a thoracic radiologist to assess its performance as a decision support tool rather than a stand-alone system.

\section*{Methods}
This study is conducted following the policy certification number 30013394 of Ethical acceptability for secondary use of medical data approved by Concordia University, Montreal, Canada. Informed consent is obtained from all the patients.

\subsection*{LDCT/ULDCT Dataset}

The LDCT/ULDCT dataset consists of volumetric Low-dose chest CT scans of 104 COVID-19 positive cases, and 56 healthy cases, collected in October 2020, December 2020, and January 2021, Babak Imaging Center, Tehran, Iran. 36.5\% of the COVID-19 cases (38 cases) are confirmed with positive RT-PCR. Diagnosis for the rest of the cases is obtained by consensus between three experienced radiologists (M.J.R., F.B.F., and A.O.) with 95.6\% agreement. The three radiologists have considered the following three main criteria when labeling the dataset:
\begin{itemize}
\item Imaging findings including GGOs, consolidation pattern, crazy paving, bilateral and multifocal lung involvement, peripheral distribution, and lower lobe predominance of findings;
\item Clinical findings including symptoms and history, and;
\item Epidemiology
\end{itemize}
The CT slices of the confirmed COVID-19 cases are then labeled by the first radiologists as having evidence of infection or not. Furthermore, all the three radiologists have scored the severity of the COVID-19 cases, by assigning a number between 1 and 4, where 1 is a mild and 4 is a severe case. The final severity score is the average over the scores from three radiologists, rounded to the nearest integer. Severity is determined based on the percentage of the lung involvement, as follows:
\begin{itemize}
\item 1: 1\%-24\%
\item 2: 25\%-49\%
\item 3: 50\%-74\%
\item 4: $\geq$75\%
\end{itemize}
Male and Female cases form 52\% and 48\% of the LDCT dataset, respectively, with the minimum age of 14 and maximum of 78. 
Male dominance is common in many COVID-19 datasets~\cite{Sun}, partly because men are more vulnerable to COVID-19~\cite{Bwire}. Furthermore, no correlation between gender and CT finding is found out~\cite{Francone}.
It is worth mentioning that to comply with the DICOM supplement 142 (Clinical Trial De-identification Profiles)~\cite{Committee2011}, we have de-identified all the CT studies.

The volumetric LDCT and ULDCT scans are obtained from a SIEMENS SOMATOM Scope scanner. All scans are in the axial view and reconstructed into $512\times512$ images using the Filtered Back Projection method~\cite{Pontana}. The radiation dose in standard chest CT scans is estimated at $7 mSv$, which is reduced to $1–1.5 mSv$ in LDCT scans and as low as $0.3 mSv$ in the ULDCT ones. For patients with $>60 kg$ body weight LDCT images are acquired using the $mAs$ value of $20$,  $kVp$ of $110 v$,  and the slice thickness of $2 mm$, whereas for patients with the body weight of less than $60 kg$ the ULDCT images are obtained with $15 mAs$.

As we did not have access to LDCT/ULDCT for CAP cases, we used a set of standard-dose volumetric chest CT scans of 60 patients~\cite{afshar:nsd}, collected before the start of pandemic from April 2018 to December 2019. This set contains 35 male and 25 female cases, with mean age of 57.7 and standard deviation of 21.7. The slices of the CAP set are also analyzed by the first radiologist to identify slices with evidence of infection. CAP images are acquired using tube current of $94-500 mA$, $kVp$ of $110-120 v$, and the slice thickness of $2 mm$, using SIEMENS SOMATOM Scope scanner.

As shown in Table~\ref{tab:data}, all cases are accompanied by demographic and clinical data, i.e., sex, age, weight, and presence or absence of 5 symptoms of cough, fever, dyspnea, chest pain, and fatigue.
We compared the performance of the proposed AI model with a fourth experienced radiologist (R.A.) who was blind to the labels, and classified the standard-dose and LDCT/ULDCT as COVID-19, CAP, and normal, first by means of the CT scans only, and then by incorporating the clinical data.

We also included an extra set of 100 positive COVID-19 patients, confirmed with positive RT-PCR. Male cases constitute 53\% of the total cases, and age average is $46.16$ with a standard deviation of $14.07$. This set was collected in April 2021.

\subsection*{Data Preprocessing}
We used a pre-trained U-Net-based lung segmentation model~\cite{Hofmanninger2020}, referred to as ``U-net (R231CovidWeb)", to segment lung regions and discard irrelevant information. This segmentation model is fine-tuned on COVID-19 images, which increases its performance and reliability for the problem at hand. Consequently, all images are downsampled from $512\times512$ to $256\times256$.

\subsection*{Two Stage Deep Learning Model}
The proposed two stage deep learning model, shown in Fig.~\ref{fig:framework}, consists of two consecutive capsule networks~\cite{Sabour, Afshar3DMCN}, which are advantageous over commonly used convolutional neural networks (CNNs) in handling the spatial relations between image instances. The segmented chest CT scans are the inputs to the first stage,  which identifies images with evidence of infection. The slice with infection could be related to a CAP or COVID-19 patient. 10 most probable slices with infection are then selected as inputs to the second stage, which consists of time-distributed capsule networks, referring to processing slices at the same time through the same model. In this stage, classification probabilities generated from individual slices go through a global max pooling operation to make the final decision. Next, the two stages are explained in more details.

\subsubsection*{Stage 1: Capsule Network}
Capsule networks are relatively new AI architectures proposed to overcome some key shortcomings of traditional deep neural networks and provide more informative features.  The key to Capsule networks' richer feature representation is the use of vectors (collection of neurons referred to as a Capsule) instead of scalars (single neurons). In other words, Capsules are groups of neurons acting as one unit, which is activated depending on the probability that a specific entity exists in the input. Capsule networks consist of layers of these Capsules stacked together to form a deep neural network and learn discriminative features from the input data.  While conventional deep learning solutions are incapable of conveying information about the relative correlations between the extracted features, Capsule networks can address this issue (via their routing by agreement mechanism) and better model existing correlations inside the network.  Through the routing by agreement process, capsules in a lower layer try to predict the output of the capsules in the next layer, and predictions are given priorities based on their correctness. The amplitude of the capsule vectors in the last layer represents the probability that the input image belongs to a specific target class. Another key advantage of Capsule networks is their ability to collect more detailed information with a smaller number of trainable parameters. This in turn results in achieving better performance with a reduced number of input data, making them the ideal AI model for the problem at hand.

The first stage of the proposed AI framework is responsible for identifying the slices demonstrating infection (caused by COVID-19 or CAP) in a series of CT images corresponding to a patient. The first stage will provide a subset of candidate slices to be analyzed in the next stage, which focuses only on the disease type. To train the first stage,  we used 2D CT images and their corresponding label (infectious vs non-infectious) to construct a slice-level classifier whose output determines the probability of the input image belonging to a specific target class (infectious vs non-infectious). We then extracted 10 slices with the highest infection probability for each patient to be used as the input of the second stage.
Given the specific characteristics of the COVID-19 disease manifestation, which include multi-focal GGOs, predominantly in peripheral, lower-lobes, and posterior anatomic areas of the lung,  we have adopted a capsule network-based classifier instead of the conventional CNN-based classifiers.  As demonstrated in our previous studies~\cite{Afshar3DMCN}, capsule networks are highly capable of capturing spatial relations between the components in medical images using small datasets and fewer parameters compared to their counterparts, which is of utmost importance in the case of COVID-19 disease.

The architecture of the first stage initiates with a stack of four convolutional layers, one pooling layer, and one batch normalization layer which are augmented by two shortcut connections to deliver shallow features to the deeper layers of the model. These layers are then followed by a stack of three capsule layers to generate the final output, which is the probability of the input image belonging to the related target class. It is worth noting that in the first stage, we are dealing with an imbalanced dataset with more number of slices without the evidence of infection. To cope with this imbalanced dataset, we have modified the loss function in the training step and considered a higher penalty for the errors in the slices demonstrating infection.

\subsubsection*{Stage 2: Time-distributed Capsule Network}
The second stage of the proposed AI framework is  a time-distributed capsule network that  takes the 10 candidates from the previous stage as inputs. These images are processed in parallel through capsule networks with the same architecture sharing all the trainable weights. These capsule networks consist of three convolutional layers, one batch normalization and one max pooling layer. The output of the last convolutional layer is reshaped to form the primary capsules, which then go through two capsule layers. The final capsule layer for each candidate corresponds to the three classes of COVID-19, CAP, and normal. To take into account the probability of the candidate slice being infected, COVID-19 and CAP classes are multiplied by the infectious probability generated by the first stage. The normal class is also multiplied by one minus the infectious probability. At the end, a global max pooling operation is applied to the outputs of the capsule networks corresponding to candidate slices, to make the final decision.

We trained the second stage time-distributed capsule network with an Adam optimizer with learning rate of $1e^{-4}$, batch size of 8, and 150 epochs. Similar to the first stage, we used a modified margin loss function to consider more penalty for the minority class. Margin loss is the original loss function for capsule networks introduced in Reference~\cite{Sabour}.

\subsection*{Incorporating the Clinical Data}
After training the two-stage deep learning model, output probabilities of the three classes (COVID-19, CAP, normal) are concatenated with the 8 clinical data (demographic and symptoms,  i.e., sex, age, weight, and presence or absence of 5 symptoms of cough, fever, dyspnea, chest pain, and fatigue) and fed to a multi-layer perceptron (MLP) model, shown in Fig.~\ref{fig:mlp}. This model has 4 fully-connected layers with 64 neurons, where each layer is followed by batch normalization. The last layer includes 3 neurons with a ``Softmax" activation function. We trained the MLP model with a cross-entropy loss and Adam optimizer with the learning rate of $1e^{-4}$, batch size of 16, and 500 epochs.

\subsection*{Grad-CAM Visualization}
We utilized the Grad-CAM localization mapping method~\cite{Selvaraju_2017_ICCV} to provide a deep insight into the intermediate layers and identify what components in a CT image have obtained the most attention by the model. The Grad-CAM method extracts the spatial information, which is preserved by the convolutional layers and specifies the parts in the image having the most contribution to the final prediction. More specifically, the Grad-CAM method generates a localization heatmap corresponding to each layer and the target class to determine the locations to which the model paid the most attention. This localization heatmap is derived by a weighted average of all feature maps in the convolutional layer followed by a Rectified Linear Unit (ReLU) activation function.

\subsection*{Statistical Analysis}
K-fold cross-validation~\cite{Nilashi2016} is a statistical approach to assess the performance of a model on an unseen dataset. According to this approach, the original dataset is randomly split into $K$ equal number of samples, and through $K$ iterations of training and testing, each of the $K$ sets is set aside for testing, and the rest is used for training. This approach is used in this study to evaluate the performance of the AI model as well as the radiologist blind to the labels, where $K$ is set to 10. Performance of each fold is reported, along with the mean and standard deviation over all the folds.
Furthermore, in each $K$ fold, 30\% of the training set is used for the validation of the associated model, according to which the most optimal model is selected and tested on the test set.
McNemar~\cite{McNemar} test, with the significance level of 0.05, is used to test the hypothesis of human and AI model having the same performance.
Logistic regression is applied to assess the significance of clinical factors in three binary classification scenarios, i.e., COVID-19 versus CAP and normal, CAP versus COVID-19 and normal, and normal versus COVID-19 and CAP.

\subsection*{Data Availability}
Data is publicly available at ``https://ieee-dataport.org/open-access/covid-19-low-dose-and-ultra-low-dose-ct-scans".

\subsection*{Code Availability}
Codes are publicly available at ``https://github.com/ParnianA/LDCT-COVID".

\subsection*{Acknowledgements}
This work was partially supported by the Natural Sciences \& Engineering Research Council (NSERC) of Canada through RGPIN-2016-04988.

\subsection*{Author Contributions}
P.A. and Sh.H. implemented the AI models. P.A., M.J.R., and A.M. drafted the manuscript. M.J.R. collected the dataset and was the first radiologist to label the patients. A.O. and F.B.F. were the two other radiologists to label the dataset. R.A. was the forth radiologist with whom the AI model was compared. F.N. and N.E. performed the statistical analysis. K.N.P. and K.F. supervised the validation of the model. A.M. supervised the study. All authors revised the manuscript.

\subsection*{Competing Interest}
The authors declare no competing interests.
\bibliography{Ref}

\begin{thebibliography}{10}
\urlstyle{rm}
\expandafter\ifx\csname url\endcsname\relax
  \def\url#1{\texttt{#1}}\fi
\expandafter\ifx\csname urlprefix\endcsname\relax\def\urlprefix{URL }\fi
\expandafter\ifx\csname doiprefix\endcsname\relax\def\doiprefix{DOI: }\fi
\providecommand{\bibinfo}[2]{#2}
\providecommand{\eprint}[2][]{\url{#2}}

\bibitem{Wu}
\bibinfo{author}{Wu, Z.} \& \bibinfo{author}{McGoogan, J.}
\newblock \bibinfo{journal}{\bibinfo{title}{Characteristics of and important
  lessons from the coronavirus disease 2019 (covid-19) outbreak in china:
  Summary of a report of 72314 cases from the chinese center for disease
  control and prevention}}.
\newblock {\emph{\JournalTitle{JAMA}}} \textbf{\bibinfo{volume}{323}},
  \bibinfo{pages}{1239--1242} (\bibinfo{year}{2020}).

\bibitem{Fang}
\bibinfo{author}{Fang, Y.} \emph{et~al.}
\newblock \bibinfo{journal}{\bibinfo{title}{Coronavirus disease 2019
  (covid-19): a systematic review of imaging findings in 919 patients}}.
\newblock {\emph{\JournalTitle{Radiology}}} \textbf{\bibinfo{volume}{296}},
  \bibinfo{pages}{E115--E117} (\bibinfo{year}{2020}).

\bibitem{Salehi}
\bibinfo{author}{Salehi, S.}, \bibinfo{author}{Abedi, A.},
  \bibinfo{author}{Balakrishnan, S.} \& \bibinfo{author}{Gholamrezanezhad, A.}
\newblock \bibinfo{journal}{\bibinfo{title}{Sensitivity of chest ct for
  covid-19: Comparison to rt-pcr}}.
\newblock {\emph{\JournalTitle{American Journal of Roentgenology}}}
  \textbf{\bibinfo{volume}{215}}, \bibinfo{pages}{87--93}
  (\bibinfo{year}{2020}).

\bibitem{Rafiee}
\bibinfo{author}{Rafiee, M.}, \bibinfo{author}{Babaki~Fard, F.},
  \bibinfo{author}{Samimi, K.}, \bibinfo{author}{Rasti, H.} \&
  \bibinfo{author}{Pressaco, J.}
\newblock \bibinfo{journal}{\bibinfo{title}{Spontaneous pneumothorax and
  pneumomediastinum as a rare complication of covid-19 pneumonia: Report of 6
  cases}}.
\newblock {\emph{\JournalTitle{Radiology case reports}}}
  \textbf{\bibinfo{volume}{16}}, \bibinfo{pages}{687--692}
  (\bibinfo{year}{2021}).

\bibitem{Miglioretti}
\bibinfo{author}{Miglioretti, D.} \emph{et~al.}
\newblock \bibinfo{journal}{\bibinfo{title}{The use of computed tomography in
  pediatrics and the associated radiation exposure and estimated cancer risk}}.
\newblock {\emph{\JournalTitle{JAMA Pediatr}}} \textbf{\bibinfo{volume}{167}},
  \bibinfo{pages}{700--707} (\bibinfo{year}{2013}).

\bibitem{Fearon}
\bibinfo{author}{Fearon, T.} \& \bibinfo{author}{Vucich, J.}
\newblock \bibinfo{journal}{\bibinfo{title}{Pediatric patient exposures from ct
  examinations: Ge ct/t 9800 scanner}}.
\newblock {\emph{\JournalTitle{AJR}}} \textbf{\bibinfo{volume}{144}},
  \bibinfo{pages}{805--809} (\bibinfo{year}{1985}).

\bibitem{Pearce}
\bibinfo{author}{Pearce, M.} \emph{et~al.}
\newblock \bibinfo{journal}{\bibinfo{title}{Radiation exposure from ct scans in
  childhood and subsequent risk of leukaemia and brain tumours: a retrospective
  cohort study}}.
\newblock {\emph{\JournalTitle{Lancet}}} \textbf{\bibinfo{volume}{380}},
  \bibinfo{pages}{499--505} (\bibinfo{year}{2012}).

\bibitem{Bell}
\bibinfo{author}{Bell, D.} \& \bibinfo{author}{Gerstmair, A.}
\newblock \bibinfo{title}{As low as reasonably achievable (alara)}.

\bibitem{Taekker}
\bibinfo{author}{Taekker, M.}, \bibinfo{author}{Kristjansdottir, B.},
  \bibinfo{author}{Ole, G.}, \bibinfo{author}{Laursen, C.} \&
  \bibinfo{author}{Pietersen, P.}
\newblock \bibinfo{journal}{\bibinfo{title}{Diagnostic accuracy of lowdose and
  ultra-low-dose ct in detection of chest pathology: a systematic review}}.
\newblock {\emph{\JournalTitle{Clinical Imaging}}}
  \textbf{\bibinfo{volume}{74}}, \bibinfo{pages}{139--148}
  (\bibinfo{year}{2021}).

\bibitem{Sakane}
\bibinfo{author}{Sakane, H.} \emph{et~al.}
\newblock \bibinfo{journal}{\bibinfo{title}{Biological effects of lowdose chest
  ct on chromosomal dna}}.
\newblock {\emph{\JournalTitle{Radiology}}} \textbf{\bibinfo{volume}{295}},
  \bibinfo{pages}{439--445} (\bibinfo{year}{2020}).

\bibitem{Park}
\bibinfo{author}{Park, J.} \emph{et~al.}
\newblock \bibinfo{journal}{\bibinfo{title}{The usefulness of low-dose ct scan
  in elderly patients with suspected acute lower respiratory infection in the
  emergency room}}.
\newblock {\emph{\JournalTitle{The British Journal of Radiology}}}
  \textbf{\bibinfo{volume}{89}} (\bibinfo{year}{2016}).

\bibitem{Schulze-Hagen}
\bibinfo{author}{Schulze-Hagen, M.} \emph{et~al.}
\newblock \bibinfo{journal}{\bibinfo{title}{Low-dose chest ct for the diagnosis
  of covid-19—a systematic, prospective comparison with pcr}}.
\newblock {\emph{\JournalTitle{Dtsch Arztebl Int}}}
  \textbf{\bibinfo{volume}{117}}, \bibinfo{pages}{389--395}
  (\bibinfo{year}{2020}).

\bibitem{Dangis}
\bibinfo{author}{Dangis, A.} \emph{et~al.}
\newblock \bibinfo{journal}{\bibinfo{title}{Accuracy and reproducibility of
  low-dose submillisievert chest ct for the diagnosis of covid-19}}.
\newblock {\emph{\JournalTitle{Radiol Cardiothorac Imaging}}}
  \textbf{\bibinfo{volume}{2}}, \bibinfo{pages}{e200196}
  (\bibinfo{year}{2020}).

\bibitem{Tofighi}
\bibinfo{author}{Tofighi, S.}, \bibinfo{author}{Najafi, S.},
  \bibinfo{author}{Johnston, S.} \& \bibinfo{author}{Gholamrezanezhad, A.}
\newblock \bibinfo{journal}{\bibinfo{title}{Low-dose ct in covid-19 outbreak:
  radiation safety, image wisely, and image gently pledge}}.
\newblock {\emph{\JournalTitle{Emergency Radiology}}}
  \textbf{\bibinfo{volume}{27}}, \bibinfo{pages}{601--605}
  (\bibinfo{year}{2020}).

\bibitem{Ardila}
\bibinfo{author}{Ardila, D.} \emph{et~al.}
\newblock \bibinfo{journal}{\bibinfo{title}{End-to-end lung cancer screening
  with three-dimensional deep learning on low-dose chest computed tomography}}.
\newblock {\emph{\JournalTitle{Nature Medicine}}}
  \textbf{\bibinfo{volume}{25}}, \bibinfo{pages}{954--961}
  (\bibinfo{year}{2019}).

\bibitem{Espinoza}
\bibinfo{author}{Espinoza, J.~L.} \& \bibinfo{author}{Dong, L.~T.}
\newblock \bibinfo{journal}{\bibinfo{title}{Artificial intelligence tools for
  refining lung cancer screening}}.
\newblock {\emph{\JournalTitle{Journal of clinical medicine}}}
  \textbf{\bibinfo{volume}{9}}, \bibinfo{pages}{3860} (\bibinfo{year}{2020}).

\bibitem{Shiri}
\bibinfo{author}{Shiri, I.} \emph{et~al.}
\newblock \bibinfo{journal}{\bibinfo{title}{Ultra-low-dose chest ct imaging of
  covid-19 patients using a deep residual neural network}}.
\newblock {\emph{\JournalTitle{European Radiology}}}
  \textbf{\bibinfo{volume}{31}}, \bibinfo{pages}{1420--1431}
  (\bibinfo{year}{2021}).

\bibitem{Sabour}
\bibinfo{author}{Sabour, S.}, \bibinfo{author}{Frosst, N.} \&
  \bibinfo{author}{Hinton, G.~E.}
\newblock \bibinfo{journal}{\bibinfo{title}{Dynamic routing between capsules}}.
\newblock {\emph{\JournalTitle{Proceedings of the 31st International Conference
  on Neural Information Processing Systems}}} \bibinfo{pages}{3859--3869}
  (\bibinfo{year}{2017}).

\bibitem{Afshar3DMCN}
\bibinfo{author}{Afshar, P.} \emph{et~al.}
\newblock \bibinfo{journal}{\bibinfo{title}{3d-mcn: A 3d multi-scale capsule
  network for lung nodule malignancy prediction.}}
\newblock {\emph{\JournalTitle{Scientific Reports}}}
  \textbf{\bibinfo{volume}{10}}, \bibinfo{pages}{7948} (\bibinfo{year}{2020}).

\bibitem{Mei}
\bibinfo{author}{Mei, X.} \emph{et~al.}
\newblock \bibinfo{journal}{\bibinfo{title}{Artificial intelligence–enabled
  rapid diagnosis of patients with covid-19}}.
\newblock {\emph{\JournalTitle{Nature Medicine}}}
  \textbf{\bibinfo{volume}{26}}, \bibinfo{pages}{1224--1228}
  (\bibinfo{year}{2020}).

\bibitem{afshar:nsd}
\bibinfo{author}{Afshar, P.} \emph{et~al.}
\newblock \bibinfo{journal}{\bibinfo{title}{Covid-ct-md, covid-19 computed
  tomography scan dataset applicable in machine learning and deep learning}}.
\newblock {\emph{\JournalTitle{Scientific Data}}} \textbf{\bibinfo{volume}{8}}
  (\bibinfo{year}{2021}).

\bibitem{McNemar}
\bibinfo{author}{McNemar, Q.}
\newblock \bibinfo{journal}{\bibinfo{title}{Psychological statistics}}.
\newblock {\emph{\JournalTitle{Wiley}}}  (\bibinfo{year}{1962}).

\bibitem{Prokop}
\bibinfo{author}{Prokop, M.} \emph{et~al.}
\newblock \bibinfo{journal}{\bibinfo{title}{Co-rads: A categorical ct
  assessment scheme for patients suspected of having covid-19—definition and
  evaluation}}.
\newblock {\emph{\JournalTitle{Radiology}}} \textbf{\bibinfo{volume}{296}},
  \bibinfo{pages}{E97--E104} (\bibinfo{year}{2020}).

\bibitem{Altmayer}
\bibinfo{author}{Altmayer, S.} \emph{et~al.}
\newblock \bibinfo{journal}{\bibinfo{title}{Comparison of the computed
  tomography findings in covid-19 and other viral pneumonia in immunocompetent
  adults: a systematic review and meta-analysis}}.
\newblock {\emph{\JournalTitle{European Radiology}}} \bibinfo{pages}{1--12}
  (\bibinfo{year}{2020}).

\bibitem{Jonas}
\bibinfo{author}{Jonas, D.} \emph{et~al.}
\newblock \bibinfo{journal}{\bibinfo{title}{Screening for lung cancer with
  low-dose computed tomography: Updated evidence report and systematic review
  for the us preventive services task force}}.
\newblock {\emph{\JournalTitle{JAMA}}} \textbf{\bibinfo{volume}{325}},
  \bibinfo{pages}{971--987} (\bibinfo{year}{2021}).

\bibitem{Armato}
\bibinfo{author}{Armato, S.} \emph{et~al.}
\newblock \bibinfo{journal}{\bibinfo{title}{The lung image database consortium
  (lidc) and image database resource initiative (idri): A completed reference
  database of lung nodules on ct scans}}.
\newblock {\emph{\JournalTitle{Medical Physics}}}
  \textbf{\bibinfo{volume}{38}}, \bibinfo{pages}{915--931}
  (\bibinfo{year}{2011}).

\bibitem{Finance}
\bibinfo{author}{Finance, J.} \emph{et~al.}
\newblock \bibinfo{journal}{\bibinfo{title}{Low dose chest ct and lung
  ultrasound for the diagnosis and management of covid-19}}.
\newblock {\emph{\JournalTitle{Journal of Clinical Medicine}}}
  \textbf{\bibinfo{volume}{10}} (\bibinfo{year}{2021}).

\bibitem{WuE}
\bibinfo{author}{Wu, E.} \emph{et~al.}
\newblock \bibinfo{journal}{\bibinfo{title}{How medical ai devices are
  evaluated: limitations and recommendations from an analysis of fda
  approvals}}.
\newblock {\emph{\JournalTitle{Nature Medicine}}}
  \textbf{\bibinfo{volume}{27}}, \bibinfo{pages}{576--584}
  (\bibinfo{year}{2021}).

\bibitem{Kleppe}
\bibinfo{author}{Kleppe, A.} \emph{et~al.}
\newblock \bibinfo{journal}{\bibinfo{title}{Designing deep learning studies in
  cancer diagnostics}}.
\newblock {\emph{\JournalTitle{Nature Reviews Cancer}}}
  \textbf{\bibinfo{volume}{21}}, \bibinfo{pages}{199--211}
  (\bibinfo{year}{2021}).

\bibitem{Sun}
\bibinfo{author}{Sun, Z.}, \bibinfo{author}{Zhang, N.}, \bibinfo{author}{Li,
  Y.} \& \bibinfo{author}{Xu, X.}
\newblock \bibinfo{journal}{\bibinfo{title}{A systematic review of chest
  imaging findings in covid-19}}.
\newblock {\emph{\JournalTitle{Quantitative imaging in medicine and surgery}}}
  \textbf{\bibinfo{volume}{10}}, \bibinfo{pages}{1058--1079}
  (\bibinfo{year}{2020}).

\bibitem{Bwire}
\bibinfo{author}{Bwire, G.}
\newblock \bibinfo{journal}{\bibinfo{title}{Coronavirus: Why men are more
  vulnerable to covid-19 than women?}}
\newblock {\emph{\JournalTitle{SN Comprehensive Clinical Medicine}}}
  \textbf{\bibinfo{volume}{2}}, \bibinfo{pages}{874--876},
  \doiprefix\url{https://doi.org/10.1007/s42399-020-00341-w}
  (\bibinfo{year}{2020}).

\bibitem{Francone}
\bibinfo{author}{Francone, M.} \emph{et~al.}
\newblock \bibinfo{journal}{\bibinfo{title}{Chest ct score in covid-19
  patients: correlation with disease severity and short-term prognosis}}.
\newblock {\emph{\JournalTitle{European Radiology}}}
  \textbf{\bibinfo{volume}{30}}, \bibinfo{pages}{6808--6817}
  (\bibinfo{year}{2020}).

\bibitem{Committee2011}
\bibinfo{author}{Committee, D.~S.}, \bibinfo{author}{Group, W.},
  \bibinfo{author}{Trials, C.} \& \bibinfo{author}{Text, F.}
\newblock \bibinfo{journal}{\bibinfo{title}{Supplement 142: Clinical trial
  de-identification profiles}}.
\newblock {\emph{\JournalTitle{DICOM Standard}}} \bibinfo{pages}{1--44}
  (\bibinfo{year}{2011}).

\bibitem{Pontana}
\bibinfo{author}{Pontana, F.} \emph{et~al.}
\newblock \bibinfo{journal}{\bibinfo{title}{Chest computed tomography using
  iterative reconstruction vs filtered back projection (part 1): evaluation of
  image noise reduction in 32 patients}}.
\newblock {\emph{\JournalTitle{European Radiology}}}
  \textbf{\bibinfo{volume}{21}}, \bibinfo{pages}{627--635}
  (\bibinfo{year}{2011}).

\bibitem{Hofmanninger2020}
\bibinfo{author}{Hofmanninger, J.} \emph{et~al.}
\newblock \bibinfo{journal}{\bibinfo{title}{Automatic lung segmentation in
  routine imaging is primarily a data diversity problem, not a methodology
  problem}}.
\newblock {\emph{\JournalTitle{European Radiology Experimental}}}
  \textbf{\bibinfo{volume}{4}}, \bibinfo{pages}{50} (\bibinfo{year}{2020}).

\bibitem{Selvaraju_2017_ICCV}
\bibinfo{author}{Selvaraju, R.~R.} \emph{et~al.}
\newblock \bibinfo{journal}{\bibinfo{title}{Grad-cam: Visual explanations from
  deep networks via gradient-based localization}}.
\newblock {\emph{\JournalTitle{Proceedings of the IEEE International Conference
  on Computer Vision (ICCV)}}}  (\bibinfo{year}{2017}).

\bibitem{Nilashi2016}
\bibinfo{author}{Nilashi, M.}, \bibinfo{author}{Ibrahim, O.} \&
  \bibinfo{author}{Ahani, A.}
\newblock \bibinfo{journal}{\bibinfo{title}{Accuracy improvement for predicting
  parkinson’s disease progression}}.
\newblock {\emph{\JournalTitle{Scientific Reports}}}
  \textbf{\bibinfo{volume}{6}} (\bibinfo{year}{2016}).

\end{thebibliography}

\end{document}